\documentclass[preprint,aps]{revtex4-1}
\pdfoutput=1

\usepackage{amsmath,amssymb,amsfonts,dcolumn,color,graphicx,graphics,latexsym,placeins,epsfig}
\usepackage{epsfig}
\usepackage{bm}
\usepackage{slashed}
\usepackage{latexsym}
\usepackage{natbib}
\usepackage{url}
\usepackage{dcolumn}
\usepackage{color}
\usepackage{amsfonts,amssymb,amsmath}
\usepackage{graphicx,epsfig}
\usepackage{psfrag}
\usepackage{subfigure}
\usepackage{tabularx}
\usepackage{hyperref}
\hypersetup{colorlinks=true}

\newcommand{\be}{\begin{equation}}
\newcommand{\ee}{\end{equation}}
\newcommand{\ba}{\begin{eqnarray}}
\newcommand{\ea}{\end{eqnarray}}

\renewcommand{\inf}{\infty}
\begin{document}

\title{Probing angle of birefringence due to long range axion hair from pulsars}
  \author{Tanmay Kumar Poddar}
\email[Email Address: ]{tanmay@prl.res.in}
\affiliation{Theoretical Physics Division, 
Physical Research Laboratory, Ahmedabad - 380009, India}
\affiliation{Discipline of Physics, Indian Institute of Technology, Gandhinagar - 382355, India}

 \author{Subhendra Mohanty}
\email[Email Address: ]{mohanty@prl.res.in}
\affiliation{Theoretical Physics Division, 
Physical Research Laboratory, Ahmedabad - 380009, India}



\begin{abstract}
Rotating neutron star or pulsar can be a possible source of pseudo Nambu Goldstone bosons or axions which can mediate long range axionic hair outside of the pulsar. When the electromagnetic radiation is emitted from pulsar and passes through the long range axion hair, the axion rotates the polarization of the electromagnetic radiation and produces birefringence. We obtain the angle of birefringence due to this long range axionic hair as $0.42^\circ$. This result is independent of the rotational frequency, radius of the pulsar, mass of the axion, and axion photon coupling constant. This value is within the accuracy of measuring the linear polarization angle of pulsar light which is $\leq 1.0^\circ$. Our result continues to hold as long as the range of the axion hair (inverse of axion mass) is greater than the radius of the pulsar, i.e; $m_a<10^{-11}$eV and the axion decay constant $f_a\lesssim \mathcal{O}(10^{17}\rm{GeV})$.
\end{abstract}


\maketitle

\section{Introduction} 
Axion was first introduced by Peccei and Quinn (1977) to solve the strong CP problem \cite{peccei,weinberg,wil,quinn}. The neutron electric dipole moment (nEDM) $d_n$ depends on a parameter $\bar{\theta}$ and from chiral perturbation theory, we can obtain $d_n\sim 10^{-16}\bar{\theta}\rm{e.cm}$. $\bar{\theta}$ is related to the QCD $\theta$ angle by $\bar{\theta }=\theta+\rm{arg\hspace{0.1cm} det} M_q$ \cite{adler,bell}, where $M_q$ is the quark mass matrix. However the current experimental bound on neutron EDM is $d_n\sim 10^{-26}\rm{e.cm}$, which implies $\bar{\theta}\lesssim 10^{-10}$ \cite{baker}. The smallness of the nEDM parameter is called the strong CP problem. To solve the strong CP problem, Peccei and Quinn postulated that $\bar{\theta}$ is not just a parameter but it is a dynamical field (scaled by $1/f_a$ to make it dimensionless) which is driven to zero by its own classical potential. It is a pseudo Nambu Goldstone boson which arises due to spontaneous breaking of global $U(1)$ symmetry at a scale $f_a$ which is called the axion decay constant and explicitly breaks at $\Lambda_{QCD}$ due to non perturbative QCD effects. These are the QCD axions which solve the strong CP problem. The QCD axion mass $(m_a)$ is related with $f_a$ by $m_a=5.7\times 10^{-12}\rm{eV}\Big(\frac{10^{18}\rm{GeV}}{f_a}\Big)$. So to obtain $f_a$ smaller than the Planck scale, the mass of the axion is $m_a\gtrsim 10^{-12}\rm{eV}$ \cite{grilli}. There are also other types of pseudo Nambu Goldstone bosons which arise due to string compactifications \cite{witten}. The detail studies of axions and axion like particles (ALPs) are discussed in \cite{wise,abbott,dine,marsh,witten}. The axions can also be a possible candidate of dark matter and even they can be ultralight \cite{hu, hui, duffy,profumo}. The coupling of axions with other particles depend on the peccei-Quinn symmetry breaking scale $f_a$. The axion can couple with photons and the coupling of axion with photons is described by the term $g_{a\gamma\gamma}a F_{\mu\nu}\tilde{F}^{\mu\nu}/4$ where $g_{a\gamma\gamma}$ is the axion photon coupling constant which is inversely proportional to $f_a$. The axion field is denoted by $a$, $F_{\mu\nu}$ is the electromagnetic field tensor and $\tilde{F}^{\mu\nu}$ is its dual.

There is no direct observational verification of axions in the universe. However one can put bound on the axion decay constant and axion mass in $m_a-f_a$ plane from laboratory, astrophysical and other experiments. There are some experiments which are searching for solar axions of mass sub-eV range and the axion decay constant $f_a\sim 10^7\rm{GeV}$ \cite{inoue,arik}. If solar axions are there in nature then it would violate supernova 1987A bound which requires $f_a\gtrsim 10^9 \rm{GeV}$. Axions can be a possible candidate of hot dark matter if $f_a\lesssim 10^8\rm{GeV}$ \cite{mirizzi,mena,hann}. Cold axions can be produced from vacuum realignment mechanism and provide a component of cold dark matter \cite{hertz,visinelli}. Axions can also be produced from the decay of axion strings \cite{shellard,kawasaki,chang}. Some laboratory experiments also put bounds on the axion parameters \cite{y,cameron,robilliard,chou,si,kim,cheng,rosen}.

The sign change in the axion potential from the coupling to gluons at large densities allows axions to be sourced by the neutron stars (NS) or pulsars \cite{hook,mairi,tanmay}. The pulsar also emits electromagnetic radiation since the dipole axis is not aligned along the spin axis. As the dipole axis precesses around the spin axis, there is a synchrotron radiation along the magnetic poles appears as the pulsed signal in a cone swept out by the radio beam. Rotating neutron star or pulsar can also form binaries with other neutron star, white dwarf and black hole and there can be ultra light scalar or pseudoscalar axion radiation \cite{panda,hook,mairi,tanmay} or vector boson radiation \cite{poddar} due to orbital period decay. Ultra light vector bosons can also mediate between the Sun and the planets and contribute to the perihelion precession of planets \cite{planet}.

Strong dipolar magnetic field induces $\textbf{E}.\textbf{B}$ density outside the pulsar which can also be a source of pseudoscalar axion and can rotate the polarization vector of the electromagnetic radiation \cite{mohanty}. There can also be galactic axion dark matter background which can rotate the polarization of the pulsar light \cite{liu}. The ALP dark matter background also rotates the CMB modes which constrains the mass of the axion and axion photon coupling constant \cite{sigl}. There is also a study where photon can change its polarization when it passes through neutrino gas \cite{nieves}. Axions can also be probed from superradiance phenomena in a polarimetric measurement for a black hole \cite{plas,shu}.

In this paper we show that pulsar can be a possible source of axions and it can produce a long range axionic hair outside of the pulsar. When the electromagnetic radiation passes through this long range axion hair it rotates the polarization of electromagnetic radiation and produces birefringence. The obtained birefringent angle for the long range axion hair is within the accuracy of measuring the linear polarization angle of pulsar light which is $\leq 1.0^\circ$ \cite{liu,crab,sigl}. The obtained birefringent angle strictly continues to hold for axions with inverse mass greater than the radius of pulsar $(10km)$ which gives $m_a<10^{-11}$eV.

The paper is organised as follows. In section II, we describe the long range axion profile for an isolated pulsar. In section III, we explain the photon propagation through the long range axion hair and calculate the birefringent angle. Finally in section IV we summarize and explain our result.

We use the units $\hslash=c=1$ throughout the paper.
\section{The long range axion profile for a pulsar }
\subsection{Effective potential of axion in vacuum}
Here we briefly discuss the result of \cite{grilli}. The interaction of axion with other standard model particles below the Peccei Quinn (PQ) and electroweak (EW) breaking scale is governed by the Lagrangian
\begin{equation}
\mathcal{L}=\frac{1}{2}\partial_\mu a\partial^\mu a+\frac{a}{f_a}\frac{\alpha_s}{8\pi}G_{\mu\nu}\tilde{G}^{\mu\nu}+\frac{1}{4}ag^0_{a\gamma\gamma}F_{\mu\nu}\tilde{F}^{\mu\nu}+\frac{\partial_\mu a}{2f_a}j^\mu_{a,0},
\label{eq:a}
\end{equation}
where the first term denotes the kinetic term of the axion field, the second term denotes the coupling of axion with the gluon field, the third term denotes the axion field couples with the photon field and the last term denotes the derivative coupling of axion field with the quark field $q$ through an axial vector current $j^\mu_{a,0}=g^0_q \bar{q}\gamma^\mu\gamma_5 q$. $\tilde{G}_{\mu\nu}=\frac{1}{2}\epsilon_{\mu\nu\sigma\delta}G^{\sigma\delta}$ is the dual of gluon field $G_{\mu\nu}$. The axion photon coupling is defined as $g^0_{a\gamma\gamma}=\frac{\alpha}{2\pi f_a}\frac{E}{N}$, where $\alpha$ is the electromagnetic fine structure constant, $E/N$ is the ratio of the electromagnetic and color anomaly respectively. Note the coupling of axion with the other standard model fields are inversely proportional to $f_a$.

We can give a chiral rotation to the quark field $q\rightarrow e^{i\gamma_5\frac{a}{f_a}Q_a}q$ so that derivative coupling disappears from the quark mass term, where $q=(u,d)$, and $\rm{tr}Q_a=1$. Hence Eq.~(\ref{eq:a}) becomes
\begin{equation}
\mathcal{L}=\frac{1}{2}\partial_\mu a\partial^\mu a+\frac{1}{4}ag_{a\gamma\gamma}F_{\mu\nu}\tilde{F}^{\mu\nu}+\frac{\partial_\mu a}{2f_a}j^\mu_a-\bar{q_L}M_aq_R+\rm{h.c}.
\label{eq:b}
\end{equation}
The new axion photon coupling constant becomes 
\begin{equation}
g_{a\gamma\gamma}=\frac{\alpha}{2\pi f_a}\Big[\frac{E}{N}-6\rm{tr}(Q_aQ^2)\Big],
\end{equation}
and the axial current density becomes
\begin{equation}
j^\mu_a=j^\mu_{a,0}-\bar{q}\gamma^\mu\gamma_5Q_aq.
\end{equation}
The quark mass matrix in the new mass basis becomes 
\begin{equation}
M_a=e^{i\frac{a}{f_a}Q_a}M_qe^{i\frac{a}{f_a}Q_a},
\end{equation}
where $M_q=\begin{bmatrix}
m_u & 0\\
0 & m_d
\end{bmatrix}$, and $Q=\begin{bmatrix}
\frac{2}{3} &0\\
0 &-\frac{1}{3}
\end{bmatrix}$. $m_u$ and $m_d$ are masses of the up and down quarks respectively.

Such chiral transformation of the quark field helps to move all the non derivative coupling into the two lightest quarks. Hence, we can integrate out all the other quarks and can work in the two flavour effective theory. $M_a$ contains non derivative couplings on the axions. So in the chiral expansion at the leading order, all the non derivative couplings of axions contain in the pion mass term governed by the Lagrangian 
\begin{equation}
\mathcal{L_\pi}\supset 2B_0\frac{f^2_\pi}{4}<U M^\dagger_a+M_a U^\dagger>,
\end{equation}
where
\begin{equation}
U=e^\frac{{i\Pi}}{f_\pi},\hspace{0.3cm} \Pi=\begin{bmatrix}
\pi^0 & \sqrt{2}\pi^{+}\\
\sqrt{2}\pi^{-} & -\pi^0
\end{bmatrix}.
\end{equation}
Here $f_\pi$ is the pion decay constant and $B_0$ is related to the chiral condensate. To derive the effective axion potential to the leading order, we only consider the neutral pion sector. Choosing $Q_a$ proportional to the identity matrix, we can write
\begin{equation}
V(a,\pi^0)=-B_0f^2_\pi\Big[m_u \cos\Big(\frac{\pi^0}{f_a}-\frac{a}{2f_a}\Big)+m_d\cos\Big(\frac{\pi^0}{f_a}+\frac{a}{2f_a}\Big)\Big],
\end{equation}
or,
\begin{equation}
V(a,\pi^0)=-m^2_\pi f^2_\pi\sqrt{1-\frac{4m_u m_d}{(m_u+m_d)^2}\sin^2\Big(\frac{a}{2f_a}\Big)}\cos\Big(\frac{\pi^0}{f_\pi}-\phi_a\Big),
\end{equation}
where 
\begin{equation}
\tan\phi_a=\frac{m_u-m_d}{m_u+m_d}\tan\Big(\frac{a}{2f_a}\Big),
\end{equation}
where $m_\pi$ is the pion mass. On the vacuum, the neutral pion gets a vev and trivially be integrated out. Hence, the effective axion potential becomes
\begin{equation}
V\approx-m^2_\pi f^2_\pi \sqrt{1-\frac{4m_um_d}{(m_u+m_d)^2}\sin^2\Big(\frac{a}{2f_a}\Big)}.
\label{eq:c}
\end{equation}
The minimum of the potential is at $<a>=0$ i.e; $\bar{\theta}=0$ and solves the strong CP problem. The second derivative of the potential gives the axion mass 
\begin{equation}
m^2_a=\frac{m_um_d}{(m_u+m_d)^2}\frac{m^2_\pi f^2_\pi}{f^2_a}.
\end{equation}
We want to probe the axions whose mass is lighter than that obtained from Eq.~(\ref{eq:c}) by a factor of $\sqrt{\epsilon}$ and we consider the parameter space $\epsilon\lesssim0.01$. For these axions, the potential becomes 
\begin{equation}
V\approx-\epsilon m^2_\pi f^2_\pi \sqrt{1-\frac{4m_um_d}{(m_u+m_d)^2}\sin^2\Big(\frac{a}{2f_a}\Big)},
\end{equation}
and consequently the mass of the axion becomes
\begin{equation}
m_a=\frac{m_\pi f_\pi}{2f_a}\sqrt{\epsilon}.
\end{equation}
\subsection{Effective axion potential at finite density}
Here we have briefly stated the result of \cite{griegel}. We are considering the astrophysical objects like neutron stars or pulsars which are made of non relativistic matter mostly like neutrons and protons. The axion potential comes from the quark mass term. From Feynman-Hellmann theorem, one can develop the in medium quark condensate. If $H(\lambda)$ denotes a Hermitian operator which depends on a real parameter $\lambda$ and operates on a normalized eigenvector $|\psi(\lambda)>$ with eigenvalue $E(\lambda)$ then 
\begin{equation}
H(\lambda)|\psi(\lambda)>=E(\lambda)|\psi(\lambda)>.
\end{equation} 
Hence, from Feynman-Hellmann theorem, we can write
\begin{equation}
<\psi(\lambda)|\frac{d}{d\lambda}H(\lambda)|\psi(\lambda)>=\frac{d}{d\lambda}<\psi(\lambda)|H(\lambda)|\psi(\lambda)>.
\end{equation}
The chiral symmetry is explicitly broken by the quark mass term which is governed by the Hamiltonian
\begin{equation}
\mathcal{H}_{mass}=m_u \bar{u}u+m_d\bar{d}d+m_s\bar{s}s+...,
\label{eq:d}
\end{equation}
where $m_u$, $m_d$, and $m_s$ denote the quark mass terms correspond to the quark fields $u$, $d$, and $s$ respectively. $...$ denotes the similar contributions from the heavier quarks. We can write Eq.~(\ref{eq:d}) as
\begin{equation}
\mathcal{H}_{mass}=2m_q\bar{q}q-\frac{1}{2}\delta m_q(\bar{u}u-\bar{d}d)+m_s\bar{s}s+...,
\end{equation}
where we define $\bar{q}q=\frac{1}{2}(\bar{u}u+\bar{d}d)$, $m_q=\frac{1}{2}(m_u+m_d)$ and $\delta m_q=m_d-m_u$. Putting $H\rightarrow \int d^3 x\mathcal{H}_{QCD}$ and $\lambda\rightarrow m_q$ in the Feynman-Hellmann theorem, we obtain
\begin{equation}
2<\psi(m_q)|\int d^3 x \bar{q}q|\psi(m_q)>=\frac{d}{d m_q}<\psi(m_q)|\int d^3 x\mathcal{H}_{QCD}|\psi(m_q)>.
\label{eq:e}
\end{equation}
Multiplying both sides by $m_q$, we obtain from Eq.~(\ref{eq:e})
\begin{equation}
2m_q<\psi(m_q)|\int d^3 x \bar{q}q|\psi(m_q)>=m_q\frac{d}{d m_q}<\psi(m_q)|\int d^3 x\mathcal{H}_{QCD}|\psi(m_q)>.
\label{eq:f}
\end{equation}
Here we neglect the isospin breaking terms, though it is not necessary. 

Consider two cases, $|\psi(m_q)>=|n_N>$ and $|\psi(m_q)>=|0>$ in Eq.~(\ref{eq:f}). $|n_N>$ denotes the ground state with nucleon density $n_N$ and $|0>$ denotes the vacuum state. Taking the differences of this two cases, we obtain
\begin{equation}
2m_q(<\bar{q}q>_{n_N}-<\bar{q}q>_0)=m_q\frac{d\xi}{dm_q}.
\label{eq:g}
\end{equation}
The nuclear matter energy density $\xi$ is given by
\begin{equation}
\xi=M_N n_N+\delta \xi,
\label{eq:h}
\end{equation}
where $\delta \xi$ contributes to the energy density from nucleon kinetic energy and nucleon nucleon interaction. The quark condensate at low density can be expressed by nucleon $\sigma$ term, $\sigma_N$ which is defined as
\begin{equation}
\sigma_N=2m_q\int d^3 x(<N|\bar{q}q|N>-<0|\bar{q}q|0>).
\end{equation}
From Eq.~(\ref{eq:f}) we obtain
\begin{equation}
\sigma_N=m_q\frac{dM_N}{dm_q}.
\label{eq:i}
\end{equation}
Combining Eq.~(\ref{eq:g}), Eq.~(\ref{eq:h}) and Eq.~(\ref{eq:i}) we obtain
\begin{equation}
2m_q(<\bar{q}q>_{n_N}-<\bar{q}q>_0)=\sigma_N n_N+...
\end{equation}
From Gell-Mann-Oakes-Renner relation, we can write
\begin{equation}
2m_q<\bar{q}q>_0=-m^2_\pi f^2_\pi.
\end{equation}
In the effective two flavour theory, we define the effective up and down quark mass as 
\begin{equation}
m^{eff}_{u,d}=m_{u,d}\Big[1-\frac{\sum \sigma^{u,d}_N n_N}{m^2_\pi f^2_\pi}\frac{m_u+m_d}{m_{u,d}}\Big].
\end{equation}
From Eq.~(\ref{eq:c}), in $m_u=m_d$ limit, the effective axion potential at finite density becomes
\begin{equation}
V=-m^2_\pi f^2_\pi \Big\{\Big(\epsilon-\frac{\sigma_N n_N}{m^2_\pi f^2_\pi}\Big) \Big|\cos\Big(\frac{a}{2f_a}\Big)\Big|+\mathcal{O} \Big(\Big(\frac{\sigma_N n_N}{m^2_\pi f^2_\pi}\Big)^2\Big)\Big\}.
\label{eq:j}
\end{equation}
\subsection{Axion profile of a pulsar}
At finite densities, the coefficient of cosine function can change sign and due to sign change in the axion potential, pulsar can be a source of axions. The mass of the axion inside the pulsar is tachyonic and it is
\begin{equation}
m_T=\frac{m_\pi f_\pi}{2f_a}\sqrt{\frac{\sigma_N n_N}{m^2_\pi f^2_\pi}-\epsilon}.
\end{equation}
Inside the pulsar, $\sigma_N\neq 0$, and $m_T\gtrsim m_a$. $\sigma_N=59\rm{MeV}$ from lattice simulation \cite{alarcon}. Axions can be a source of pulsar if the size of the pulsar is larger than a critical size $r_c$, where 
\begin{equation}
r_c\gtrsim \frac{1}{m_T}.
\label{eq:k}
\end{equation}
A typical pulsar of mass $M=1.4 M_{\odot}$ and radius $R=10\rm{km}$ can be a source of axions if Eq.~(\ref{eq:k}) is satisfied which gives the upper bound on the axion decay constant, $f_a\lesssim 2.352\times 10^{17}\rm{GeV}$.

The axion potential has degenerate vacua and this degeneracy can be broken by higher dimensional operators at planck scale or finite density effect \cite{rr}. Outside of the pulsar, the potential attains minima at $a=0,\pm 4\pi f_a,...$ and maxima at $a=\pm 2\pi f_a, \pm 6\pi f_a...$ etc. Inside of the pulsar, the potential has maxima at $a=0,\pm 4\pi f_a,...$ and minima at $a=\pm 2\pi f_a, \pm 6\pi f_a...$ etc.

Inside the pulsar, the axion field is tachyonic and reside on one of the local maxima of the axion potential. Throughout inside of the pulsar, the dynamical axion field takes a constant value $a=4\pi f_a$ at the nearest local maximum. Outside of the pulsar, the axion field rolls down to the nearest local minimum and asymptotically reaches $a=0$ value at infinity.

For an isolated pulsar of constant density, the axion field equation of motion becomes \cite{tanmay,hook}
\begin{equation}
\nabla^{\mu}\nabla_{\mu} \left(\frac{a}{2f_a}\right)= \begin{cases} 
     -m_T^2\sin \left(\frac{a}{2f_a}\right) \text{sgn}\lbrace \cos \left(\frac{a}{2f_a}\right)\rbrace & (r<R),\\
       m_a^2\sin \left(\frac{\theta}{2}\right) \text{sgn}\lbrace \cos \left(\frac{a}{2f_a}\right)\rbrace & (r>R).
   \end{cases}
\end{equation}
 
\begin{figure}[!htbp]
\centering
\subfigure[$V$ vs. $a$]{\includegraphics[width=3.0in,angle=360]{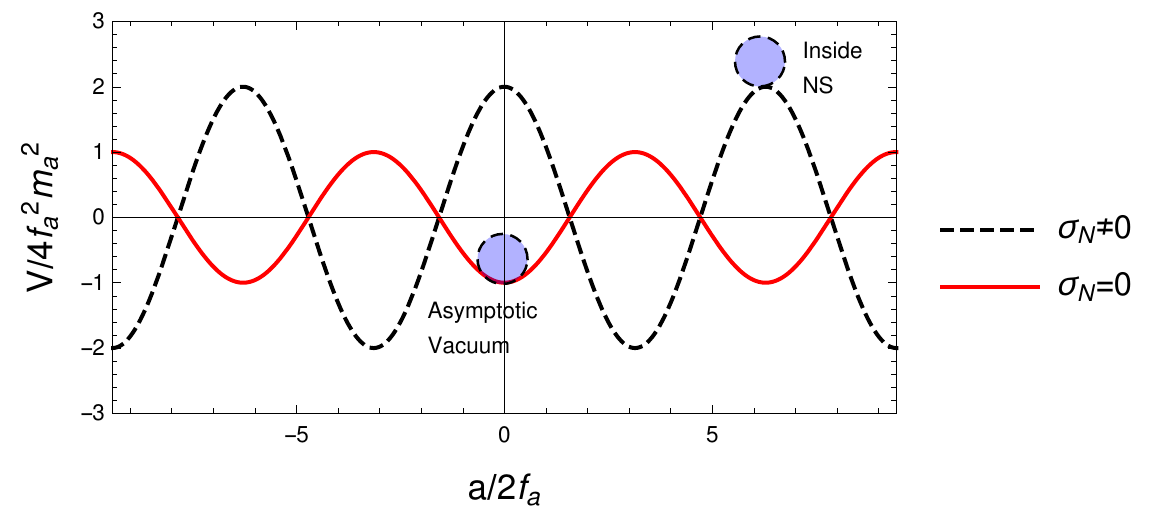}\label{subfig:vvsa}}
\subfigure[$V$ vs. $r$]{\includegraphics[width=2.5in,angle=360]{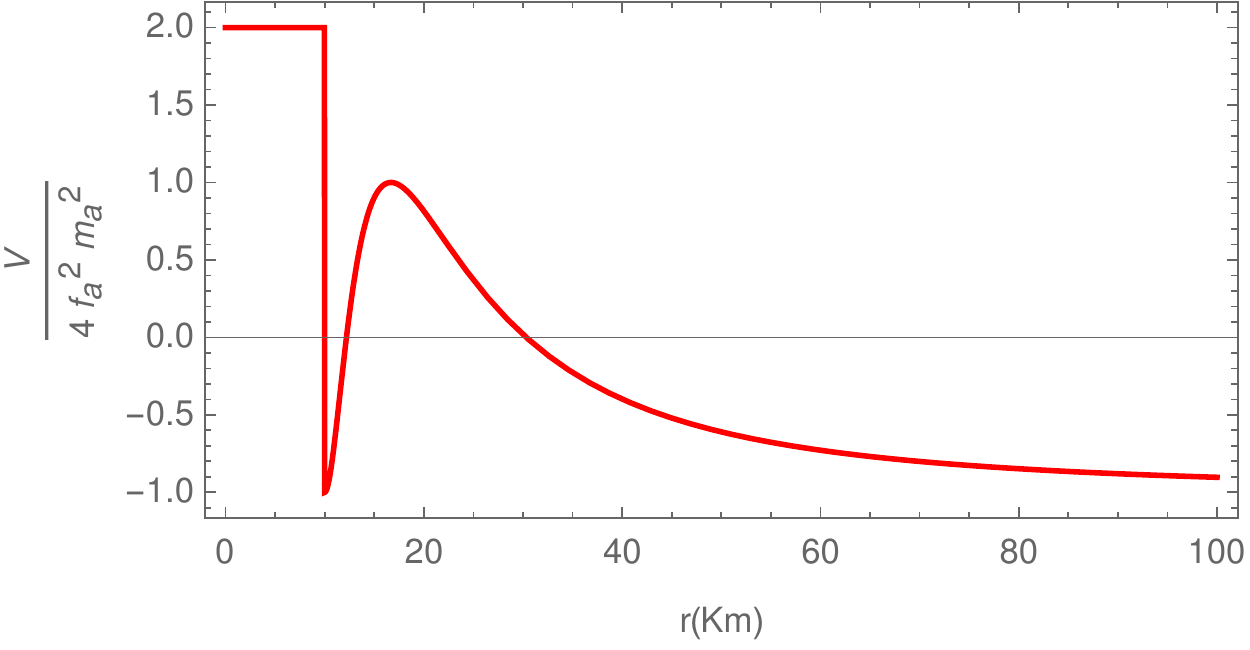}\label{subfig:vvsr}}
\subfigure[$a$ vs. $r$]{\includegraphics[width=3.0in,angle=360]{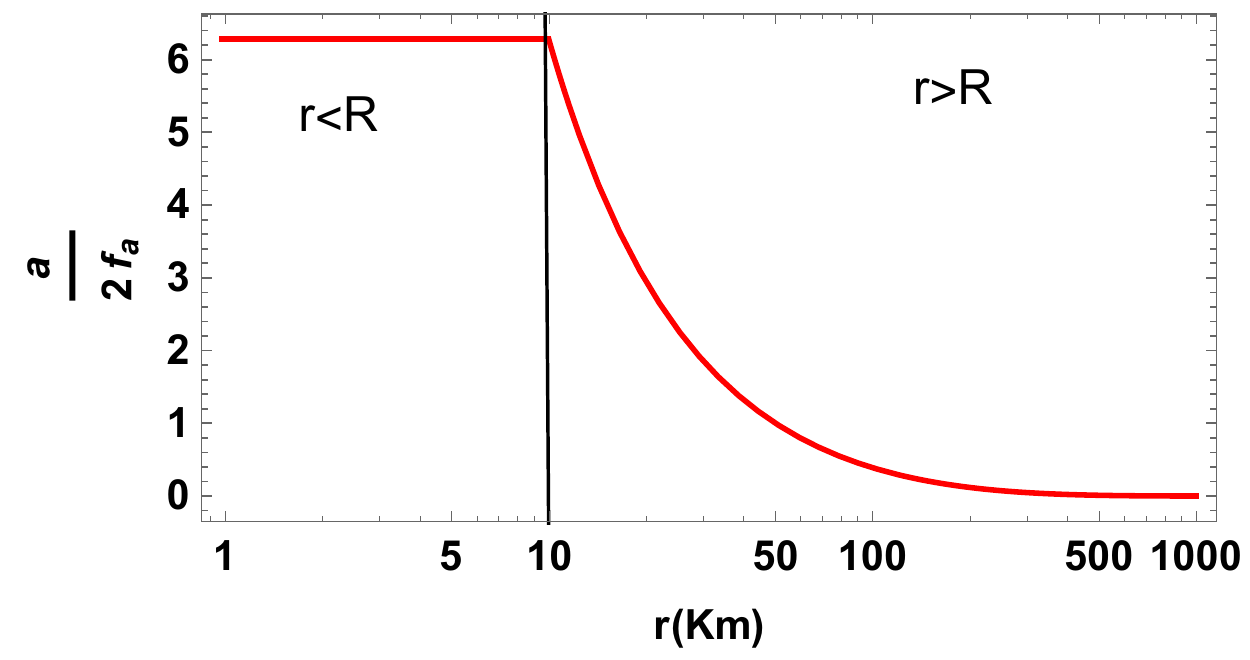}\label{subfig:avsr}}
\subfigure[$q$ vs. $m_a$]{\includegraphics[width=3.0in,angle=360]{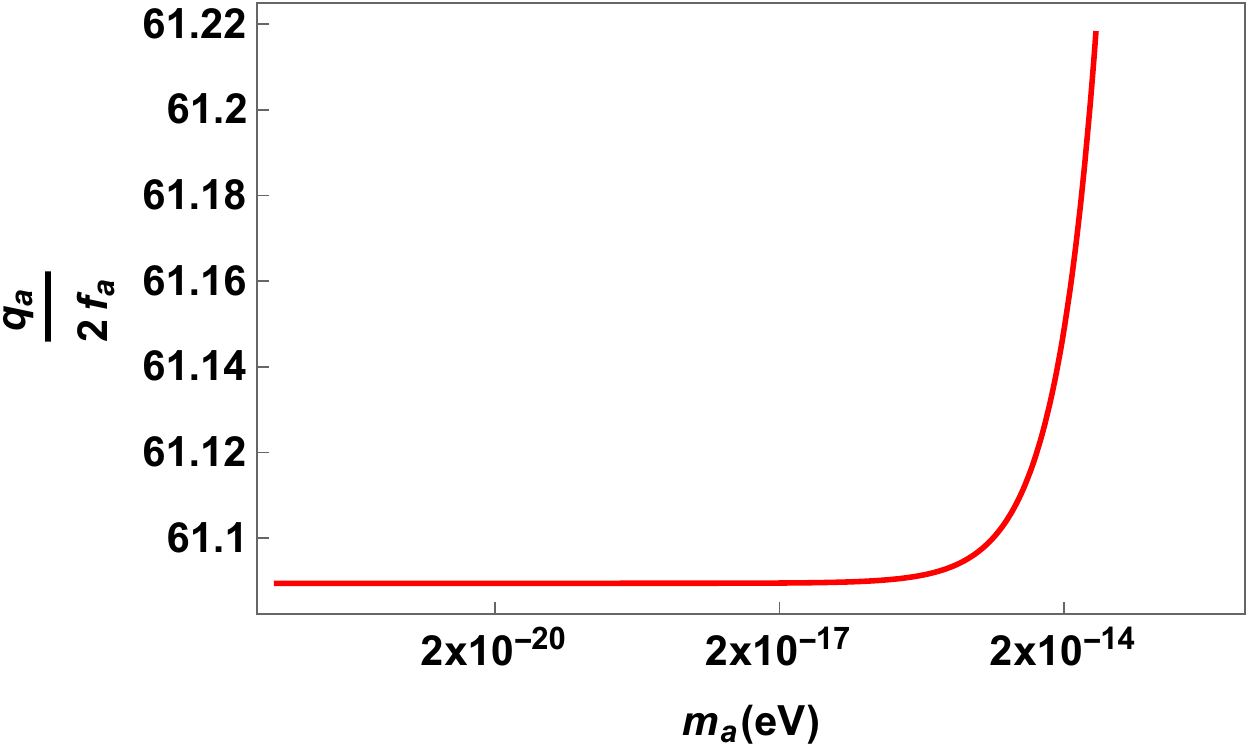}\label{subfig:avsr}}
\caption{(a)plot of axion potential V with respect to $a$. We take $m_T^2/m_a^2=2$. The black dashed line corresponds to inside the pulsar and the red solid line corresponds to outside the pulsar. (b) Plot of $V$ as a function of distance $(r)$ from the centre of pulsar. There is a discontinuity of $V$ at $R$ due to sign change in $V$. (c)Plot of $a$ as a function of $r$.(d)Variation of charge with axion mass.}
\label{fig:axion_profile}
\end{figure}

If a neutron star or pulsar is immersed in a low mass axionic potential then the solution of the axion field outside of the pulsar falls off with distance like Yukawa interaction. Assuming the spacetime outside the pulsar is Schwarzschild, the axion field equation of motion becomes
\begin{equation}
\left(1-\frac{2GM}{r}\right)\frac{d^2a}{dr^2}+\frac{2}{r}\left(1-\frac{GM}{r}\right)\frac{da}{dr}=m_a^2a,
\label{eq:dd}
\end{equation}
where $M$ is the mass of the pulsar, and $G$ is the universal gravitational constant. In the Schwarzschild background, the axion field will not satisfy the Klein-Gordon equation. So we expand the axion field in a perturbative way where the perturbation parameter is $GM/R$. Hence, the axion field solution becomes \cite{tanmay,planet}
\begin{equation}
a(r)=\frac{q_ae^{-m_ar}}{r}\Big[1+\frac{GM}{Rr}\{1-m_ar\ln(m_ar)+m_are^{2m_ar}E_i(-2m_ar)\}\Big]+\mathcal{O}\Big(\Big(\frac{GM}{R}\Big)^2\Big),
\label{eq:ee}
\end{equation}
where $q_a$ is the effective axion charge of the pulsar. From the continuity of axion field at the surface of the pulsar, we obtain the effective axion charge
\begin{equation}
q_a=4\pi f_a R e^{m_aR}\Big[1+\frac{GM}{R^2}\{1-m_a R \ln(m_aR)+m_aRe^{2m_aR}E_i(-2m_aR)\}\Big]^{-1}+\mathcal{O}\Big(\Big(\frac{GM}{R}\Big)^{-2}\Big)
\label{eq:ff}
\end{equation}
Hence, the axion field solutions inside and outside of the pulsar are 
\begin{eqnarray}
a&=&\frac{q_a e^{-m_ar}}{r}\Big[1+\frac{GM}{Rr}\{1-m_ar\ln(m_ar)+m_are^{2m_ar}E_i(-2m_ar)\}\Big]+\mathcal{O}\Big(\Big(\frac{GM}{R}\Big)^2\Big),
r>R.\nonumber\\
&=&4\pi f_a, \hspace{11.0cm} \hspace{1.0cm}r<R.
\label{eq:kk}
\end{eqnarray}
We assume that the spacetime metric outside the pulsar is Schwarzchild because for a typical pulsar $(M=1.4M_\odot, R=10Km)$, $GM/R$ is not very much small. The plot of axion field and axion potential inside and outside of a pulsar and the variation of axion charge with axion mass is shown in FIG.\ref{fig:axion_profile} \cite{tanmay}. 
The space time outside of a rotating neutron star or pulsar is described by the approximate form of the Kerr metric,
\begin{equation}
ds^2=\Big(1-\frac{2GM}{r}\Big)dt^2-\Big(1-\frac{2GM}{r}\Big)^{-1}dr^2-r^2(d\theta^2+\sin^2\theta d\phi^2)-\frac{4GMj}{r}\sin^2\theta  d\phi dt.
\end{equation} 
Where $j$ is the angular momentum per unit mass and this approximation is valid as long as $j\ll GM$. For a typical pulsar $j=10^{-6}km$, $GM=1.5km$, and we can neglect the last term. Hence the metric outside of the pulsar effectively becomes Schwarzchild \cite{mohanty}.
\section{Photon propagation in an axionic field: Birefringence}
When a linearly polarized pulsar light passes through a long range axionic hair which is originated from the same pulsar then due to their dispersion relations the left and the right circular polarization modes will attain opposite corrections. This effect is called birefringence.

The Lagrangian which describes the interaction between the axion couples with two photons is
\begin{equation}
\mathcal{L}=\frac{1}{2}(\partial_\mu a)(\partial^\mu a)-\frac{1}{4}F^{\mu\nu}F_{\mu\nu}-\frac{1}{4}g_{a\gamma\gamma}aF_{\mu\nu}\tilde{F}^{\mu\nu}.
\end{equation}
Here $g_{a\gamma\gamma}$ is given as
\begin{equation}
g_{a\gamma\gamma}=\frac{c\alpha_{em}}{2\pi f_a},
\label{eq:cy}
\end{equation}
where $c$ is a model dependent parameter of order unity and $\alpha_{em}$ is the electromagnetic fine structure constant. The modified Maxwell's field equations in presence of axion coupling become\cite{mohanty}
\begin{equation}
\nabla.\textbf{E}=-g_{a\gamma\gamma}(\nabla a).\textbf{B},
\label{eq:one}
\end{equation}
\begin{equation}
\nabla\times \textbf{B}-\frac{\partial \textbf{E}}{\partial t}=g_{a\gamma\gamma}\Big[(\nabla a)\times\textbf{E}+\textbf{B}\frac{\partial a}{\partial t}\Big],
\label{eq:two}
\end{equation} 
\begin{equation}
\nabla.\textbf{B}=0,
\label{eq:three}
\end{equation}
\begin{equation}
\nabla\times \textbf{E}+\frac{\partial \textbf{B}}{\partial t}=0,
\label{eq:four}
\end{equation}
where $\textbf{E}$ and $\textbf{B}$ are the elctric and magnetic field vectors of the electromagnetic radiation. Propagation of the electromagnetic radiation from pulsar through the long range axionic hair is governed by combining the above equations Eq.~(\ref{eq:one}), Eq.~(\ref{eq:two}), Eq.~(\ref{eq:three}) and Eq.~(\ref{eq:four}):
\begin{equation}
\nabla_\mu\nabla^\mu\textbf{B}=-g_{a\gamma\gamma}(\nabla a)\times\frac{\partial\textbf{B}}{\partial t}.
\label{eq:five}
\end{equation}
Suppose this magnetic field has a harmonic variation and we can write
\begin{equation}
B(x,t)=\mathcal{B}e^{i\phi(x,t)},
\end{equation} 
where $\mathcal{B}$ is the magnetic field of electromagnetic radiation and $\phi$ is the phase. A linearly polarized wave is an equal admixture of right and left circularly polarized waves. So we can write the transverse magnetic field as
\begin{equation}
\mathcal{B}_{\pm}=\mathcal{B}_i\pm i\mathcal{B}_j,
\end{equation}
where $\mathcal{B}_i$ and $\mathcal{B}_j$ are the components of $\mathcal{B}$ along $\hat{e}_i$ and $\hat{e}_j$ orthogonal to $\hat{r}$. In the Fourier space we can write Eq.~(\ref{eq:five}) as
\begin{equation}
k_\mu k^\mu\mathcal{B}=ig_{a\gamma\gamma}(\nabla a\times\omega \mathcal{B}),
\label{eq:six}
\end{equation}
where $k^\mu=(\omega,\textbf{k})$ is the photon four momentum.  Eq.~(\ref{eq:six}) decouples for the two circularly polarized modes and we obtain
\begin{equation}
k_\mu k^\mu \mathcal{B}_\pm \mp g_{a\gamma\gamma}(\partial_r a)\omega\mathcal{B}_\pm=0.
\end{equation}
Hence the dispersion relation for the two circularly polarized modes due to different phase velocities propagating radially from the poles of the pulsar is \cite{mohanty}
\begin{equation}
\omega^2\Big(1-\frac{2GM}{r}\Big)^{-1}-k^2_r\Big(1-\frac{2GM}{r}\Big)=\pm g_{a\gamma\gamma}(\partial_r a)\omega.
\label{eq:seven}
\end{equation}
Hence we can write the propagation vector from Eq.~(\ref{eq:seven}) as
\begin{equation}
k_r=\omega\Big(1-\frac{2GM}{r}\Big)^{-1}\mp \frac{g_{a\gamma\gamma}}{2}(\partial_r a).
\label{eq:eight}
\end{equation}
So the phase shift between the left and the right circularly polarized modes is
\begin{equation}
\Delta\phi=\int^\inf_R (k^+_r-k^-_r)dr.
\end{equation}
Using Eq.~(\ref{eq:eight}) we can rewrite the phase shift as
\begin{equation}
\Delta\phi=g_{a\gamma\gamma}[a(\inf)-a(R)].
\label{eq:en}
\end{equation}
The axion field has a long range behaviour outside of the pulsar. Using Eq.~(\ref{eq:kk}), Eq.~(\ref{eq:cy}), and Eq.~(\ref{eq:en}) we can write the phase difference as
\begin{equation}
\Delta\phi=-\frac{c\alpha_{em}}{2\pi f_a}\frac{q_a e^{-m_aR}}{R}\Big[1+\frac{GM}{R^2}\{1-m_aR\ln(m_aR)+m_aRe^{2m_aR}E_i(-2m_aR)\}\Big], 
\label{eq:pola}
\end{equation}
where the axion field goes to zero at infinity. Positive sign of phase shift implies anti clockwise rotation and negative sign of phase shift implies clockwise rotation by looking down the light path.

The observed birefringent angle which is the angle of rotation of the linear polarization $(\Delta\theta)$ is half the phase shift $\Delta\phi$ between the two circular polarization modes. Using Eqs.~(\ref{eq:ff}), and Eqs.~(\ref{eq:pola}) we obtain
\begin{equation}
\Delta\theta=-c\alpha_{em}, 
\label{eq:nine}
\end{equation}
where, $c$ is a model dependent parameter of $\mathcal{O}(1)$ and $\alpha_{em}=1/137$. Hence we obtain the birefringent angle 
\begin{equation}
\Delta\theta=7.299\times 10^{-3}radian=0.42^{\circ}
\label{eq:ten}
\end{equation}

\begin{figure}
\centering
\includegraphics[width=4.0in,angle=360]{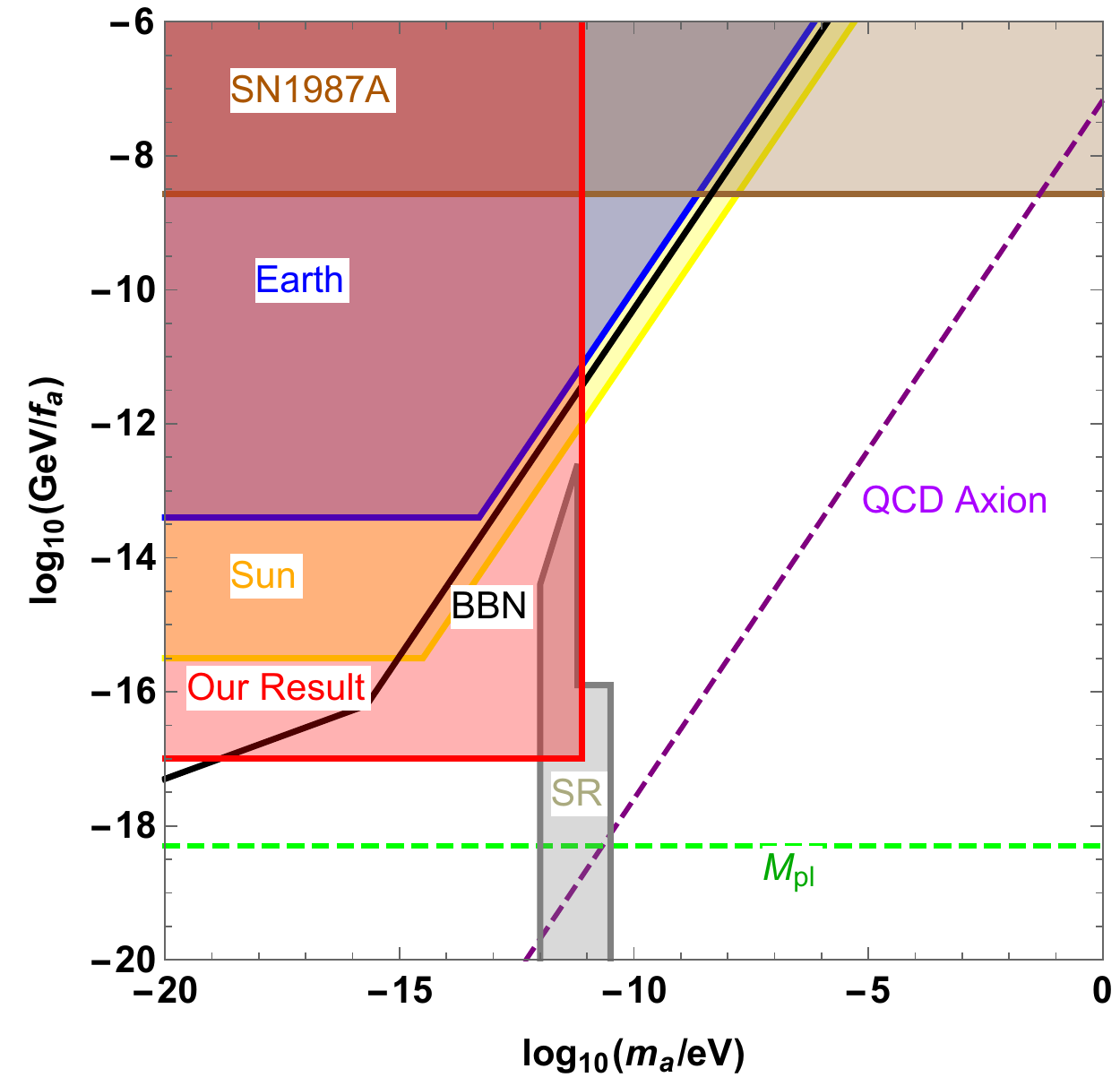}\label{subfig:vvsa}
\caption{The brown shaded region is excluded by SN1987A, the blue shaded region is excluded by direct measurement of earth, and the yellow shaded region is excluded by direct measurement of Sun. The gray shaded region is excluded by blackhole superradiance measurements \cite{super,radiance}. The green dotted line denotes the reduced planck scale. The black continuous line denotes the constraints from BBN if axion is the dark matter \cite{bbn}. The violet dotted line corresponds to QCD axions. Our result of the birefringent angle can probe the red shaded region.}
\label{fig:axion_p}
\end{figure}

External magnetic field can also give rise such type of rotation of the polarization vector which is called the Faraday effect. The main difference between the optical rotation by long range axionic hair and Faraday effect is that for Faraday effect the birefringent angle is proportional to $\lambda^2$ where $\lambda$ is the wavelength of the electromagnetic wave and for our case i.e; the optical rotation by axion field, the birefringent angle is independent of $\lambda$. The existing constraints on axion parameters and the region in which our result is valid are shown in FIG.\ref{fig:axion_p} 
\section{Discussions}
We have discussed if axions are the source of pulsar and if it mediates a long range axionic hair outside, then the axion hair can rotate the polarization of the electromagnetic radiation of the pulsar. Here we do not need any external magnetic field for the optical rotation. The birefringent angle that we have first derived is independent of the angular frequency of rotation, radius of the pulsar, mass of the axion, and the axion photon coupling constant. Our result is true for axions of mass $m_a<10^{-11}$eV and $f_a\lesssim \mathcal{O}(10^{17}\rm{GeV})$. We obtain the birefringent angle as $0.42^{\circ}$ if the pulsar has a long range axionic hair. The derived birefringent angle is within the accuracy of measuring the linear polarization angle of pulsar light which is $\leq 1.0^\circ$.

\end{document}